\documentclass[prd,eqsecnum,floatfix]{revtex4}%
\usepackage{graphicx}
\usepackage{amsfonts}
\usepackage{amssymb}%

\newcommand{\be}{\begin{equation}}
\newcommand{\ee}{\end{equation}}
\newcommand{\bea}{\begin{eqnarray}}
\newcommand{\eea}{\end{eqnarray}}

\newcommand{\nn}{\nonumber\\}

\begin{document}

\title{Neutrino oscillations in a stochastic model for space-time foam}

\author{J. Alexandre}
\affiliation{Kings's College London, Department of Physics, London WC2R 2LS, U.K.}
\email{jean.alexandre@kcl.ac.uk}

\author{K. Farakos}
\affiliation{Department of Physics, National Technical University of Athens, Zografou Campus, 157 80 Athens, Greece}
\email{kfarakos@central.ntua.gr}

\author{N. E. Mavromatos}
\affiliation{Kings's College London,~Department of Physics, London WC2R 2LS, U.K.}
\email{nikolaos.mavromatos@kcl.ac.uk}

\author{P. Pasipoularides}
\affiliation{Department of Physics,~National Technical University of Athens, Zografou Campus, 157 80 Athens, Greece}
\email{paul@central.ntua.gr}

\begin{abstract}
We study decoherence models for flavour oscillations in four-dimensional stochastically fluctuating space times and discuss briefly the sensitivity of current neutrino experiments to such models. We pay emphasis on demonstrating the model dependence of the associated decoherence-induced damping coefficients in front of the oscillatory terms in the respective transition probabilities between flavours. Within the context of specific models of foam, involving point-like D-branes and leading to decoherence-induced damping which is inversely proportional to the neutrino energies, we also argue that future limits on the relevant decoherence parameters coming from TeV astrophysical neutrinos, to be observed in ICE-CUBE, are not far from theoretically expected values with Planck mass suppression. Ultra high energy neutrinos from Gamma Ray Bursts at cosmological distances can also exhibit in principle sensitivity to such effects.
\end{abstract}

\maketitle
\section{Introduction\label{sec:1}}

The possibility that quantum gravity involve models with stochastic
fluctuations of the associated metric field, around some fixed
background value, say flat Minkowski space time, may not be an
unrealistic one. Such stochastic models of gravity lead to
observable consequences in principle, ranging from light cone
fluctuations~\cite{lcf} to induced decoherence for matter
propagating in such fluctuating space times~\cite{sarkar}. Space
time foam models  naturally include such stochastic space time
backgrounds.

We describe here briefly the possible appearance of a stochastic
space-time background by the D-particle foam model. It should be
remarked that foam models do not consist exclusively of microscopic
black holes, although this is the most popular model. The authors of
\cite{horizons} have provided another type of foam, inspired by
brane-world scenarios. According to this model of \emph{D-particle
foam}, our brane world moves in a bulk space, punctured by
point-like D-particles, which are point-like defects in membrane
theory, characterised though by an infinity of super-Planckian
string states. During such a motion, D-particles from the bulk cross
our brane world, and interact with propagating matter in it,
represented by open strings with their end-points attached on the
brane. The interactions of string matter with the D-particle defects
may not be universal among particle species. Due to electric charge
conservation, electrically charged states, such as quarks or
electrons may undergo only smooth ordinary scattering (recoil),
while electrically neutral states, such as photons or neutrinos, may
undergo topologically non trivial interactions in which a string
state is split into two, with (some of) the corresponding end points
are attached to the D-particle (``capture''). Capture includes
recoil of the defect,  as well as the induced back reaction onto
space time, determined by means of conformal field theory methods of
the associated non-critical string corresponding to the capture
process~\cite{horizons}. For heavy (Galilean) D-particles, the
analysis leads to induced metric deformations of the (initially)
Minkowski, say, flat background of the form:
\begin{equation}
g_{00}=-1, g_{ij}=\delta_{ij}, \quad g_{i 0} = g_{0 i} = \frac{v_i }{c}
\label{metric}
\end{equation}
where $i,j$ spatial (four-dimensional)  indices on the brane world,
with $0$ denoting a temporal direction, and $v_i/c$ is the recoil
3-velocity of the D-particle defect, in units of the speed of light
in vacuum, $c$, which by means of momentum conservation during the
scattering process equals
\begin{equation}\label{velmag}
v_i = g_s \Delta k_i/M_s ~.
\end{equation}
In the last relation, $\Delta k_i$  denotes the pertinent momentum
transfer of matter, and $M_s/g_s$ is the D-particle mass, with $M_s$
the string scale and $g_s < 1$ the (weak) string
coupling~\cite{horizons,recoil}. We stress once more that the form
(\ref{metric}) is valid for Galilean (heavy) D-particles, relative
to the momenta of the incident matter states. In general, for
relativistic D-particles there are higher order corrections of $v_i$
in the expression for the recoil-induced metric, leading to a
covariant form in terms of the \emph{four-velocity } $u_\mu = \gamma
(1, {\vec v})$, with $\gamma $ the appropriate Lorentz
factor~\cite{sakellariadou}:
 \begin{equation}
g_{\mu\nu}= \eta_{\mu\nu} + f(\Phi)u_\mu u_\nu
\label{metriccov}
\end{equation}
where the scalar factor $f(\Phi)$ depends on the details of the
scalar dilaton $\Phi$ configuration in the model. For instance, in
the impulse approximation to D-particle recoil, the $\Phi$ field is
proportional to $u_\mu X^\mu $, as dictated by (logarithmic)
conformal field
theory considerations~\cite{sakellariadou}, and $f(\Phi) =
\Theta_\epsilon (u_\mu X^\mu )$, with $\epsilon \to 0^+$ an
appropriate regulator, linked to the world-sheet scale on account of
closure of the conformal algebra~\cite{recoil}. However, more
general forms for the function $f(\Phi)$ are possible, depending on
the model.

As discussed in \cite{sarkar}, the above mentioned capture process
may entail flavour oscillation as well as a non-trivial Fock vacuum
structure of the vacuum experienced by the flavour states instead of
the corresponding mass eigenstates. When considering statistical
distributions of particles, one may have on average $\langle u_i
\rangle = 0$, with only non trivial stochastic fluctuations $\langle
u_i u_j \rangle = \sigma^2 \delta_{ij}$, $\sigma^2 \ne 0$. The
distribution of such D-particle velocities is a model dependent
concept. In view of (\ref{metric}), a stochastic statistical
fluctuation of $u_i$ will also result in a stochastic fluctuation of
the induced metric, and thus a stochastically fluctuating gravity
theory.

In \cite{sarkar} we have studied the implications of such stochastic
metrics in the simplified (but not unrealistic) case where the
D-particle recoil velocities where along the direction of motion of
the particle probe, thereby leading to an effectively two space-time
dimensional problem, in which, however, the $\gamma$-matrix
structures associated with the flavoured fermions (neutrinos) were
kept four dimensional. For definiteness, in that work we have
considered a model of Gaussian fluctuations, and we demonstrated
that the associated oscillation probabilities of both fermion and
boson flavoured particles exhibited exponential damping and
decoherence, with the damping exponent scaling with the square of
the time variable $t$.

It is the purpose of this work to extend these results to more
generic models of stochastically fluctuating metrics, living in a
full-fledged four dimensional space time. In particular, we assume
small random fluctuations $h^{\mu\nu}$, around the Minkowski
background inverse metric $\eta^{\mu\nu}$,
\begin{equation}
g^{\mu\nu}=\eta^{\mu\nu}+h^{\mu\nu} \label{fluc}
\end{equation}
where for $\eta_{\mu\nu}$ we assume the sign convention
$(1,-1,-1,-1)$ and $h^{\mu\nu}$ are defined as stochastic variables,
obeying some statistical distribution, defining the model, such that
$\langle h_{\mu\nu} \rangle = 0~, $ but $\langle
h_{\mu\nu}h_{\rho\sigma}\rangle$ are non trivial.

The effect of neutrino oscillations will be studied in the above
mentioned stochastic quantum gravity environment (\ref{fluc}).
Measurable quantities, such as transition amplitudes between
neutrino flavours, will be obtained by averaging over the stochastic
variables $h^{\mu\nu}$. As we will see, the influence of the quantum
gravity environment has a decoherence effect on the neutrino
states, which is equivalent to a damping of the oscillations between
neutrino flavours.

The determination of the particular statistical distribution for the
random fluctuations $h^{\mu\nu}$ requires compete knowledge of the
quantum gravity theory. In the absence of such a theory, the
statistical distribution can not be fixed. Hence, our considerations
below will be largely phenomenological, in an attempt to impose
bounds on the relevant parameters by comparison to experiment. To
this end, we mention that the allowed distributions must be such
that the corresponding characteristic function yields finite,
well-defined results for physical observables, such as the flavour
oscillation probability. This criterion allows the use of
distributions, such as the Cauchy-Lorentz (Breit-Wigner resonance
type), for which the mean (or odd moments) may not be well defined,
while their variance (and higher even moments) diverge by
themselves. For other distributions, such as the
Gaussian~\cite{sarkar}, of course the latter entities are well
defined and small. It is these two kinds of distributions for the
metric fluctuations that we shall concentrate upon in this work.
Within the context of our D-particle foam model~\cite{horizons},
such distributions may characterise the ensembles of the D-particle
recoil velocities. As we will see in the following sections, the
choice of the distribution leads to different damping in the
oscillation probability. For instance, the Cauchy-Lorentz case is
characterised by exponential damping, with a linear power of time
$t$, as in the Lindblad decoherence model~\cite{lindblad,benatti},
in contrast to the Gaussian model~\cite{sarkar}, where the damping
exponents are proportional to $t^2$.

In this work we shall be only concerned with stochastic gravity
models, with fluctuations (\ref{fluc}) about flat Minkowski
space-time backgrounds. In such models we saw that the results for
the flavour oscillation probabilities are identical between fermions
and bosons. The situation may be different, if the background space
times are not Minkowski flat. We shall come back to such more
general situations in a forthcoming publication. Comparison of these
two kinds, insofar as the order of magnitude of the expected
effects, and their prospects of discovery in future experimental
facilities, will be discussed in the concluding section of our
article.

The structure of the article is as follows: in section \ref{sec:2}
we set up the formalism for scalar flavoured particles,
concentrating on the case of two generations for simplicity. We
examine first the situation, in which only leading order (linear)
terms in the deformation $h_{\mu\nu}$ are kept in the expressions
for the pertinent oscillation probabilities, paying attention to
demonstrate the dependence on the results, in particular the form of
the damping exponents as functions of time, on the kind of the
(statistical) distribution of the metric fluctuations. We also
demonstrate that the inclusion of higher order corrections in
$h_{\mu\nu}$ do not change the results qualitatively. In section
\ref{sec:3} we extend the analysis to two-generations of fermions,
showing that the results concerning the oscillation probabilities
and the effects of decoherence due to the stochastic metric
fluctuations, are identical to the scalar case. Finally we comment
briefly on a possible extension of the above results to the
realistic three generation neutrino case, which however shall be
considered in detail in a forthcoming publication. Conclusions and
outlook are presented in section \ref{sec:4}.

\section{Flavour Oscillations for Scalar particles in a stochastic metric background \label{sec:2}}

In this section we study massive scalar particles in the quantum
gravity stochastic background (\ref{fluc}). The Klein-Gordon
equation for a single scalar particle in a gravitational background,
reads:
\begin{equation}\label{equaboson}
\left( g^{\alpha\beta} \nabla_{\alpha}\nabla_{\beta}+m^2 \right) \phi=0
\end{equation}
where $g^{\alpha\beta}$ is the inverse metric tensor and
$\nabla_{\alpha}$ is a gravitational covariant derivative.

A flat fluctuation $h^{\mu\nu}$ around the inverse Minkowski metric
$\eta^{\mu\nu}$ is defined by Eq. (\ref{fluc}). For simplicity, and
motivated by the Galilean D-particle foam case (\ref{metric}), we
may assumed that $h^{\mu\nu}$ is independent of space-time
coordinates~\footnote{This is not the case for the fully
relativistic situation (\ref{metriccov}), which may exhibit
coordinate dependence through the dilaton field $\Phi (x)$, or in
general for cases of inhomogeneous velocities.}. In such a case, the
Christoffel symbols vanish, hence the Klein-Gordon equation in 3+1
dimensions reads:
\begin{equation}\label{kg}
\left(g^{00} \partial_{0}\partial_{0}+ 2 g^{0i}\partial_{0}\partial_{i}+g^{ij} \partial_{i}\partial_{j}+m^2\right)
\phi=0
\end{equation}
We seek for plane wave solutions of the form:
\begin{equation}
\phi(x,t)=\hat\phi(k,\omega)e^{i (\omega t-k x)} \label{pl1}
\end{equation}
where $k$ represents the momentum of the scalar particle in the
Minkowski metric background. Note, that we have chosen one of the
coordinate axes to lie along the direction of motion of the scalar
particles, i.e. we have set $k_2=k_3=0$ and $k_1=k$. Plugging the
expression (\ref{pl1}) into eq.(\ref{equaboson}) leads to
\begin{equation}
g^{00}\omega^{2}-2 g^{01}k \; \omega+g^{11} k^2-m^2=0 \label{dr1}
\end{equation}
The two solutions $\omega (k)$, $\omega'(k)$ of the above equation
are physically equivalent, since
\begin{equation}
\omega (-k)=-\omega'(k)~.
\end{equation}
Keeping the positive energy solution, we obtain the dispersion
relation
\begin{equation}
\omega=\frac{g^{01}}{g^{00}} k+\frac{1}{g^{00}}\sqrt{(g^{01})^2 k^2
-g^{00}(g^{11}k^2-m^2)} \label{dr}
\end{equation}
We next proceed to discuss the flavour oscillations, by considering
first the approximation of keeping at most linear terms in the
stochastic fluctuations $h_{\mu\nu}$.

\subsection{Linear approximation for the stochastic metric fluctuations}

We assume for brevity two flavours for the scalar particles, written
as a two-component scalar $\Phi=(\phi_1,\phi_2)$. The equation of
motion is \be \left( g^{\alpha\beta}
\nabla_{\alpha}\nabla_{\beta}+M^2 \right) \Phi=0 \ee where $M^2$ is
the $2\times 2$ positive definite (mass)$^2$ matrix, with
eigenvalues $m^2_1$ and $m^2_2$. Note, that the flavour states are
not eigenstates of $M$. For $t=0$, we assume the production of a
scalar particle of flavour $\alpha$ with  density matrix:
\begin{equation}
\rho(0)= |\phi_{\alpha}\rangle\langle \phi_{\alpha}| \label{rho0}
\end{equation}
Taking into account that the energy eigenstates $|f_{l}\rangle$ ($l=1,2$) are related to
the flavour states $|\phi_{\alpha}\rangle$ via the unitary transformation:
\begin{eqnarray}
&&|\phi_{\alpha}\rangle= \sum_{i} U_{\alpha i} |f_{i}\rangle \quad  \quad \langle\phi_{\alpha}|=\sum_{j} U^{*}_{\alpha j} \langle f_{j}|\label{orth0} \\
&&|f_{i}\rangle=\sum_{\beta}U^{*}_{\beta i}|\phi_{\beta}\rangle  \quad  \quad \langle f_{j}|=\sum_{\gamma} U_{\gamma j} \langle \phi_{\gamma}| \label{orth}
\end{eqnarray}
we then obtain from Eqs. (\ref{rho0}) and (\ref{orth0}):
\begin{equation}
\rho(0)= \sum_{i,j}U_{\alpha i}U^{*}_{\alpha j} |f_{i}\rangle \langle f_{j}| \label{rho1}
\end{equation}
where we sum only over $i,j$, with $\alpha$ denoting the initial flavour state.
The time evolution of the density matrix yields:
\begin{equation}
\rho(t)= \sum_{i,j}U_{\alpha i}U^{*}_{\alpha j} e^{i(\omega_{i}-\omega_{j})t} |f_{i}\rangle \langle f_{j}| \label{rho2}
\end{equation}
Upon using Eq. (\ref{orth}), one obtains the density matrix at time t, expressed in the basis of flavour states:
\begin{equation}
\rho(t)=\sum_{i,j,\beta,\gamma}U_{\alpha i}U^{*}_{\alpha j} U^{*}_{\beta i} U_{\gamma j} e^{i (\omega_{i}-\omega_{j})t} |\phi_{\beta}\rangle\langle \phi_{\gamma}|~.
\end{equation}
The probability to find the scalar particle in a state of flavour $\beta$ ($\beta\neq \alpha$) is
\begin{equation}
\mbox{Prob}(\alpha \rightarrow \beta)=\sum_{i,j} U_{\alpha i}  U^{*}_{\alpha j} U_{\beta i}^{*} U_{\beta j} e^{i (\omega_{i}-\omega_{j})t} \label{prob}
\end{equation}
where the time dependent part is
\begin{equation}
U_{\alpha 1}  U^{*}_{\alpha 2} U_{\beta 1}^{*} U_{\beta 2}  e^{i (\omega_{1}-\omega_{2})t}+ U_{\alpha 2}  U^{*}_{\alpha 1} U_{\beta 2}^{*} U_{\beta 1} e^{i (\omega_{2}-\omega_{1})t}
\end{equation}
Since the perturbations $h_{\mu\nu}$ are stochastic variables, it is necessary to compute the following integral
\begin{equation}\label{mean1}
\langle e^{i (\omega_{1}-\omega_{2})t} \rangle=\int d^{10}{\bf h} ~F(\textbf{h}) e^{i (\omega_{1}-\omega_{2})t}
\end{equation}
where $\textbf{h}=(h^{00}, h^{01}, h^{02},...,h^{23},h^{33})$, and $F(\textbf{h})$ is a multi-variable probability density function.
The measurable transition probability is found by averaging over the metrics
\begin{equation}
\langle \mbox{Prob}(\alpha \rightarrow \beta)\rangle=\sum_{i,j} U_{\alpha i}  U^{*}_{\alpha j} U_{\beta i}^{*} U_{\beta j} \langle e^{i (\omega_{i}-\omega_{j})t} \rangle \label{prob1}
\end{equation}
In what follows we will assume that the stochastic variables $h^{\mu\nu}$ are independent, or equivalently that:
\begin{equation}
 F(\textbf{h})=f(h^{00}) f(h^{01})f(h^{02})\times... \times f(h^{23}) f(h^{33})
\end{equation}
where $f(x)$ is a one-variable probability density function for the stochastic variable $x$. Additionally, we have assumed that the one-variable   density function $f(x)$ is the same for all $h^{\mu\nu}$.
We will approximate the energy difference $\omega_1-\omega_2$ with a linear expansion of Eq. (\ref{dr}) over the fluctuations $\textbf{h}$
\begin{equation}\label{exp0}
\omega_1(\textbf{h})-\omega_2(\textbf{h})=a-\frac{a}{2} h^{00}-\frac{b}{2} h^{11}+O(\textbf{h}^2)
\end{equation}
where
\begin{eqnarray}\label{def1}
a=\sqrt{k^2+m_1^2}-\sqrt{k^2+m_2^2}, ~~~~~~ b=\frac{k^2}{\sqrt{k^2+m_1^2}}-\frac{k^2}{\sqrt{k^2+m_2^2}}
\end{eqnarray}
Note that in the above linear expansion only the elements $h^{00}$ and $h^{11}$ are involved.
The integral (\ref{mean1}) then reduces to
\bea\label{mean12}
\langle e^{i (\omega_1-\omega_2)t} \rangle&=&\int dh^{00}dh^{11} ~f(h^{00})~f(h^{11})~
\exp\left\lbrace i t\left( a-ah^{00}/2-bh^{11}/2\right)\right\rbrace\nn
&=&\Phi\left(-at/2\right)\Phi\left(-bt/2\right) \exp(iat)
\eea
where
\begin{equation}
\Phi(\xi)=\int_{-\infty}^{+\infty}e^{i \xi x}f(x)dx
\end{equation}
is the characteristic function of the stochastic variable $x$ with probability density function $f(x)$.

\subsection{Gaussian fluctuations}

We will assume first a Gaussian distribution for the metric fluctuations, with density function
\begin{equation}
f(x)=\frac{e^{-x^2/\sigma^2}}{\sqrt{\pi \sigma^2}}
\end{equation}
where we have considered the case of zero mean value ($\mu=0$), and $\sigma$ is the corresponding standard deviation.
The characteristic function for the Gaussian distribution is:
\begin{equation}\label{cfgd}
\Phi(\xi)=\frac{1}{\sqrt{\pi \sigma^2}} \int_{-\infty}^{+\infty}dx ~
\exp\left( i \xi x-x^2/\sigma^2\right) =\exp\left(-\xi^2 \sigma^2/4\right)
\end{equation}
From Eq. (\ref{mean12}) we obtain
\begin{equation}\label{mean2}
\langle e^{i (\omega_{1}-\omega_{2})t} \rangle=
\exp\left\lbrace iat-\frac{\sigma^2 t^2}{16} \left(a^2+b^2\right)\right\rbrace \label{Gaussian}
\end{equation}
Hence, the amplitude of the flavour oscillations has an exponential
damping quadratic in time, that dependents on the particle's
momentum. In the case of high energy particles (compared to their
mass) one can make an asymptotic expansion for $m_i/k<<1$
\begin{equation}
\langle e^{i (\omega_{1}-\omega_{2})t} \rangle \simeq \exp\left\lbrace ikt\Delta \left(1-\frac{m_1^2+m_2^2}{4 k^2}\right) \right\rbrace
\exp\left\lbrace -\frac{\sigma^2 (kt)^2}{8}\Delta^2 \right\rbrace\label{Gaussian1}
\end{equation}
where terms of order ${\mathcal O}(m_i/k)^6$ have been disregarded, and
the parameter $\Delta$ is defined as
\be
\Delta=\frac{m_1^2-m_2^2}{2k^2}<<1.
\ee
We remark at this point that self-consistency in the expansion in powers of $m_i/k$ forces us to keep the forth-order correction term in the expression for the oscillation frequency in (\ref{Gaussian1}), since the damping is of order $(m_i/k)^4$. The results are in full agreement with the corresponding effective two-dimensional case of \cite{sarkar}. A discussion on the prospects of observing such a decoherence effect in realistic physical systems of neutrinos, as well as on the dependence of the effect on the distance of the neutrino sources, especially in the case of astrophysical neutrinos, will be given in the concluding section of the article.

\subsection{Cauchy-Lorentz fluctuations}

We will consider next the case of Cauchy-Lorentz fluctuations
\begin{equation}
f(x)=\frac{1}{\pi}\frac{\gamma}{x^2+\gamma^2}
\end{equation}
where $\gamma$ is the scale parameter, and the location parameter is zero.
Note that the Cauchy-Lorentz distribution has no well defined mean, and in general odd higher-order moments, while
its variance and higher-order even moments diverge. However, the characteristic function, defined as:
\begin{equation}\label{cfcld}
\Phi(\xi)=\frac{1}{\pi} \int_{-\infty}^{\infty}dx~\frac{\gamma~e^{i \xi x}}{x^2+\gamma^2} =\exp\left( -\gamma |\xi|\right)
\end{equation}
yields finite results for the oscillation probabilities:
\begin{equation}\label{mean3}
\langle e^{i (\omega_{1}-\omega_{2})t} \rangle=\exp\left\lbrace i a t- \frac{\gamma t}{2}(|a|+|b|)\right\}
\end{equation}
In this situation we obtain an exponential damping, with the exponent linear in time. An expansion in powers of $m_i/k$ gives
\begin{equation}
\langle e^{i (\omega_{1}-\omega_{2})t} \rangle\simeq
\exp\Big\{ ikt\Delta-\gamma kt |\Delta|\Big\} \label{Cauchy}.
\end{equation}
where terms of order ${\mathcal O}(m_i/k)^4$ have been disregarded.
We did not write here the correction to the oscillation frequency, as this is of order $(m_i/k)^4$,
whereas the damping is of order $(m_i/k)^2$, and we are interested in the dominant effect only.
The case is similar in form to the Lindblad decoherence~\cite{lindblad,benatti}.

\subsection{Beyond the linear approximation}

In this section we go beyond the linear approximation for Gaussian metric fluctuations, in order to check whether the previously derived damping of flavour oscillations in the linear-$h$ approximation is modified significantly. As we shall see, the main results remain valid.
In this case we expand the energy difference $\omega_1-\omega_2$ up to quadratic terms in the fluctuations $\textbf{h}$
\begin{equation}
\omega_1(\textbf{h})-\omega_2(\textbf{h})=a+{\bf h \cdot d} +{\bf h \cdot D \cdot h}
+ O(\textbf{h}^3)
\end{equation}
with ${\bf h}=(h^{00},h^{01},h^{11})$, ${\bf d}=(-a/2 ,0, -b/2)$ and
\begin{eqnarray}
\textbf{D}=\left(\begin{array}{ccc}
  3a/4 &0 & b/4 \\
 0 & b & 0 \\
 b/4  & 0 & -c/4
\end{array}\right)
\end{eqnarray}
The parameters $a$ and $b$ are given by Eq. (\ref{def1}), while $c$ is defined as:
\begin{equation}
c=\frac{k^4}{k^2+m_1^2}-\frac{k^4}{k^2+m_2^2}
\end{equation}
We next write the Gaussian probability density for the metric fluctuations in a compact form:
\begin{equation}
F({\bf h})=\frac{e^{-{\bf h\cdot\Xi\cdot h}}}{(\sqrt{\pi}\sigma)^3}
\end{equation}
where
\begin{equation}
{\bf \Xi}=\mbox{diag}\left(\frac{1}{\sigma^2},\frac{1}{\sigma^2},\frac{1}{\sigma^2}\right)
\end{equation}
We wish to compute the integral
\begin{equation}\label{mean6}
\langle e^{i (\omega_{1}-\omega_{2})t} \rangle=\int d^{3}{\bf h} ~
\frac{e^{-{\bf h\cdot\Xi\cdot h}}}{(\sqrt{\pi}\sigma)^3} e^{i (\omega_{1}-\omega_{2})t}
\end{equation}
using the formula
\begin{equation}
\int d^{3}{\bf h} ~\exp\left( -{\bf h\cdot B\cdot h}+{\bf u\cdot h}\right)
=\frac{\pi^{3/2}}{\mbox{det}{\bf B}} \exp\left( \frac{\bf u \cdot B^{-1}\cdot u}{4} \right) \label{form1}
\end{equation}
where the matrices $\textbf{B}$ and $\textbf{u}$ are defined as:
\begin{eqnarray}
&&\textbf{B}={\bf\Xi}-i t \textbf{D}\\
&&{\textbf{u}}=it \textbf{d}
\end{eqnarray}
The matrix $\textbf{B}$ can be written explicitly:
\begin{eqnarray}
\textbf{B}=\left(\begin{array}{ccc}
  \frac{1}{\sigma^2}-\frac{3}{4} i ta &0 & -\frac{1}{4} it b \\
 0 &  \frac{1}{\sigma^2}-i t b & 0 \\
-\frac{1}{4} it  b  & 0 &  \frac{1}{\sigma^2}+\frac{1}{4}i t  c
\end{array}\right)
\end{eqnarray}
Upon applying Eq. (\ref{form1}), we find:
\begin{eqnarray}
\langle e^{i (\omega_{1}-\omega_{2})t} \rangle&=&
\left(\frac{\mbox{det}{\bf\Xi}}{\mbox{det}{\bf B}}\right)^{1/2}\exp(-\chi(t)) \exp(i a t) \nonumber \\
&=&\frac{4}{\sqrt{P(t)}}\exp(-\chi(t))~\exp(i a t)
\end{eqnarray}
where
\begin{eqnarray}
\chi(t)&=&\frac{\sigma^2 t^2 \left(4(a^2+b^2)-4 i \sigma^2 t (b^2-ac)\right)}{4\sigma^4 t^2(b^2+3 a c)-16 i \sigma^2 t (3 a-c)+64}\nn
P(t)&=&(1-i b\sigma^2 t)(16-4 i(3a-c)\sigma^2 t+(b^2+3 a c)\sigma^4 t^2)
\end{eqnarray}
Expanding the exponent $\chi(t)$ in powers of the small parameter $\sigma$ we obtain
\bea
\chi(t)&=&\frac{\sigma^2 t^2}{16} (a^2+b^2)+\frac{i\sigma^4 t^3}{64} (3 a^3+2 a b^2-b^2 c)\nn
&&-\frac{\sigma^6 t^4}{256} (9 a^4+7 a^2 b^2+b^4-2 a b^2 c+b^2 c^2)+O(\sigma^8)
\eea
The leading term of the above expansion corresponds to the main effect which is the exponential damping of particle oscillations. Note that it is identical with the one found in the linear approximation above. The next to leading order term is purely imaginary and modifies slightly the oscillation term $e^{i a t}$. The factor $P(t)$ has a subleading contribution of the form:
\begin{equation}
|P(t)|=16+\frac{\sigma^4 t^2}{2}(9 a^2+18 b^2+c^2)+O(\sigma^6)
\end{equation}
The reader should compare the results with the effectively two-dimensional model of \cite{sarkar}.
The results are similar, independently of the representation of gravity fluctuations and the number of spatial dimensions.

\section{Fermions in a stochastic metric background \label{sec:3}}

In this section we would like to extend the above results to
incorporate particles with spin (fermions), which from a
phenomenological point of view is more interesting, in view of the
potential application to the physics of neutrino oscillations.
However, as we shall demonstrate below, the results will be similar
to the bosonic case, as far as the main features of decoherence
damping is concerned. This is attributed to expanding about a flat
Minkowski background (\ref{fluc}).

\subsection{Dirac fermion dispersion relation}

In this section, we will consider Dirac fermions, but the results
presented are the same for Majorana fermions, as the only difference
is that the latter are described by self-conjugate fields, which
does not affect the dispersion relation. We remark that this
difference would play a r\^ole only if we considered the MSW
effect~\cite{MSW}, for the propagation of neutrinos in matter media,
as in that case only one Weyl spinor component of the fermion would
be affected. We shall consider this case, along with more general
situations involving expansions about non flat backgrounds in a
future publication.

We review here the basic elements we will need to describe fermions
in a curved background. We shall follow the formalism of
\cite{borde} and in references therein, where we refer the reader
for details. At each point ${\bf M}$ of space time, the vierbeins
${\bf e}_\alpha=\partial_\alpha {\bf M}$ span the flat tangent space
time, and are related to the inverse metric by \be e^\mu_\alpha~
e^\nu_\beta~\eta^{\alpha\beta}=g^{\mu\nu} \ee The gamma matrices
$\gamma^\alpha$ which follow are defined on the flat tangent space
time and satisfy
$\{\gamma^\alpha,\gamma^\beta\}=2\eta^{\alpha\beta}$, and we also
define
$\sigma^{\alpha\beta}=i[\gamma^\alpha,\gamma^\beta]/2$.\\
The Dirac equation in a curved background is
\be
\left( i\gamma^\alpha{\mathcal D}_\alpha-m\right) \psi=0
\ee
where
\be
{\mathcal D}_\alpha=e^\mu_\alpha\left[ \partial_\mu-\frac{i}{4}e^\nu_\beta\nabla_\mu e_{\rho\nu}\sigma^{\beta\rho}\right]
\ee
and $\nabla_\mu$ is the covariant derivative.
In the present situation, we are interested in a constant and homogeneous stochastic metric, such that the vierbeins
do not depend on space time coordinates, and the Christoffel symbols vanish. As a consequence, the Dirac equation reads
\be\label{dirac}
\left( i\gamma^\alpha e_\alpha^\mu\partial_\mu-m\right) \psi=0
\ee
In order to find the dispersion relation for the fermion, we multiply the Dirac equation (\ref{dirac})
by the complex conjugate operator $ (-i\gamma^\beta e_\beta^\nu\partial_\nu-m)$ to obtain
\bea
0&=&\left( \frac{1}{2}\left\lbrace \gamma^\alpha,\gamma^\beta\right\rbrace
 e_\alpha^\mu ~ e_\beta^\nu\partial_\mu\partial_\nu+m^2\right) \psi\nn
&=&\left(g^{\mu\nu}\partial_\mu\partial_\nu+m^2\right) \psi
\label{disferm} \eea which is similar to the equation
(\ref{equaboson}) for a boson, since covariant derivatives are
simple derivatives in our case. As a consequence, the spin does not
play a role in the situation where the background metric is flat,
and the previous results derived for a boson apply to fermions. It
appears that damping is a general result of quantum gravity
fluctuations.

\subsection{Two-flavour fermion oscillations}

We consider two Dirac fermions $\psi_e,\psi_\mu$, written as an
eight-component fermion $\Psi$, which are coupled by a Dirac mixing
mass matrix, and are described by the equation of motion
\be
\left(i\gamma^\alpha{\mathcal D}_\alpha-M\right) \Psi=0 \label{dirac2} ,
\ee
where the mass matrix in flavour space reads:
\be M=\left(
\begin{array}{cc}
m_e & m_{e\mu} \\
m_{e\mu} & m_\mu\\
\end{array}\right) \label{unitary}
\ee
In order to involve the mass eigenstates $\psi_1,\psi_2$, one performs the following rotation in flavour
space \cite{mannheim}
\be\label{rotation}
\left(\begin{array}{c} \psi_e \\ \psi_\mu \end{array}\right) =
\left(\begin{array}{cc} \cos\theta & \sin\theta \\ -\sin\theta & \cos\theta
\end{array}\right)
\left(\begin{array}{c} \psi_1 \\ \psi_2 \end{array}\right)
\ee
where
\be
\tan(2\theta)=\frac{2m_{e\mu}}{m_\mu-m_e}
\ee
As explained in \cite{mannheim},
the Lagrangian describing the fermions $\psi_1$ and $\psi_2$ is then the sum of two
independent Lagrangians, such that the equations of motion for $\psi_1$ and $\psi_2$ are
\be
\left( i\gamma^\alpha{\mathcal D}_\alpha-m_j\right)\psi_j=0~~~~~~~~j=1,2,
\ee
where $m_1+m_2=m_e+m_\mu$ and $m_1m_2=m_em_\mu-m_{e\mu}^2$.
Hence, this leads to the study of two individual fermions.
The results are then similar to those derived for bosons, as the fermion dispersion relation is identical with that of bosons, as can be readily seen from Eq. (\ref{disferm}).

Thus, using Eq. (\ref{prob}) and Eqs. (\ref{Gaussian}), (\ref{Cauchy}) and (\ref{rotation}), one obtains the formula for the transition probability between flavours, in the linear-order approximation for the $h$-metric fluctuations:
\begin{equation}\label{proba2}
\langle \mbox{Prob}(\alpha\rightarrow \beta)\rangle=\frac{1}{2} \sin^2 (2 \theta) \left(1-  e^{-\chi(t)} \cos( a t)\right),
\end{equation}
where $\alpha\ne\beta$ and
\begin{itemize}
\item $\chi(t)=\frac{\sigma^2 t^2}{16} \left(a^2+b^2\right)$ for the Gaussian case;
\item $\chi(t)=\frac{\gamma t}{2}(|a|+|b|)$ for the Cauchy-Lorentz case.
\end{itemize}
We note that, in the case where there is no metric fluctuation
(for $\chi(t)=0$), the probability (\ref{proba2}) leads to the known result
$\frac{1}{2}\sin^2(2\theta)[1-\cos(at)]$ \cite{henley}. Taking into
account metric fluctuations, we see that the limit $t\to\infty$ leads to the stationary
probability $\frac{1}{2}\sin^2(2\theta)$, which is independent of the fermion energy.

\subsection{Three-flavour fermion oscillations}

It is not difficult to generalize our results in the case
of three flavours. As we will see the exponential damping remains.

The corresponding $3\times3$ Dirac mass matrix $M$, defined in analogy with Eq. (\ref{unitary}), can be diagonalized with a unitary transformation $U_{\alpha i}$:
\begin{equation}
|\psi_{\alpha}\rangle=\sum_i U_{\alpha i}| \psi_{i} \rangle \quad \quad (\alpha=e,\mu,\tau)
\end{equation}
and has three positive eigenvalues $m_{i} \quad (i=1,2,3)$. We note for completeness that for the U matrix we use the parametrization of \cite{henley}.

The probability to find a fermion, with initial state $\alpha$, in a final state with flavour $\beta\ne\alpha$ is
given by Eq. (\ref{prob1}):
\begin{equation}\label{proba3}
\langle\mbox{Prob}(\alpha \rightarrow \beta)\rangle=\sum_{i,j} U_{\alpha i}  U^{*}_{\alpha j} U_{\beta i}^{*} U_{\beta j} \langle e^{i (\omega_{i}-\omega_{j})t}\rangle.
\end{equation}
As in the two-flavour case,
the probability (\ref{proba3}) has a time-independent part, which is
\begin{equation}
\mid U_{\alpha 1}\mid^2 \mid U_{\beta 1}\mid^2+\mid U_{\alpha 2}\mid^2 \mid U_{\beta 2}\mid^2+\mid U_{\alpha 3}\mid^2 \mid U_{\beta 3}\mid^2,
\end{equation}
whereas its time-dependend part is
 \begin{equation}
 U_{\alpha 1}  U^{*}_{\alpha 2} U_{\beta 1}^{*} U_{\beta 2} \langle e^{i (\omega_{1}-\omega_{2})t}\rangle+ U_{\alpha 2}  U^{*}_{\alpha 3} U_{\beta 2}^{*} U_{\beta 3} \langle e^{i (\omega_{2}-\omega_{3})t}\rangle+U_{\alpha 3}  U^{*}_{\alpha 1} U_{\beta 3}^{*} U_{\beta 1} \langle e^{i (\omega_{3}-\omega_{1})t}\rangle+cc.
\end{equation}
We then average over the stochastic fluctuations $h^{\mu\nu}$, and calculate
 the three elements
\begin{center}
$\langle e^{i (\omega_{1}-\omega_{2})t} \rangle,~~~~~~\langle e^{i (\omega_{2}-\omega_{3})t} \rangle,~~ ~~~~
\langle e^{i (\omega_{3}-\omega_{1})t} \rangle$.
\end{center}
Using our previous results (see Eqs. (\ref{Gaussian1}) and (\ref{Cauchy})), we obtain for Gaussian fluctuations
\begin{equation}\label{gauss2}
\langle e^{i (\omega_{i}-\omega_{j})t} \rangle \simeq \exp\left\lbrace ikt\Delta_{ij} \left(1-\frac{m_i^2+m_j^2}{4 k^2}\right) \right\rbrace
\exp\left\lbrace -\frac{\sigma^2 (kt)^2}{8}\Delta_{ij}^2 \right\rbrace,
\end{equation}
and for Cauchy-Lorentz fluctuations
\begin{equation}\label{cl}
\langle e^{i (\omega_{i}-\omega_{j})t} \rangle \simeq
\exp\Big\{ ikt\Delta_{ij}-\gamma kt|\Delta_{ij}|\Big\} ,
\end{equation}
where
\begin{equation}\label{ddef}
\Delta_{ij}=\frac{m_i^2-m_j^2}{2k^2}<<1.
\end{equation}
The reader is reminded at this point that, as with (\ref{Gaussian1}), consistency in the expansion in powers of $m_i/k$  forces us to keep the
forth-order correction in $m_i/k$ in the oscillation frequency (\ref{gauss2}) of the Gaussian model, given that the damping is of order $(m_i/k)^4$.
On the other hand, in the Cauchy-Lorentz case (\ref{cl}) we disregarded such a forth order correction, since  the corresponding damping is of order $(m_i/k)^2$.

We now notice that the $1/E$-dependent exponential damping in the case (\ref{cl}) may be translated~\cite{barger,icecube} as implying a finite life time $\tau_{\rm lab}$ for the probe in the laboratory frame:
\begin{equation}\label{lifetime}
       {\rm exp}\left(-t~\frac{\gamma |m_i^2 - m_j^2|}{2E}\right) \equiv
        {\rm exp}\left(-\frac{t}{\tau_{\rm lab}}\right) =  {\rm exp}\left(-\frac{t m_{\nu_i}}{E_{\nu_i}\tau_{\nu_i}}\right)
\end{equation}
where $\tau_{\nu_i}$ is the life-time of the neutrino probe in the
rest frame of the massive neutrino. It is interesting therefore to
use our microscopic models in order to identify a possible
quantum-gravitational origin of a finite lifetime of neutrinos. We
shall come back briefly to this point later on the article, when we
discuss lower bounds on the decoherence-induced neutrino lifetime
(\ref{lifetime}) from (current and future) experiments.

\section{Conclusions and Outlook \label{sec:4}}

In this concluding section we would like to discuss briefly the
above results in light of the prospects for observing the above
effects in current or upcoming neutrino facilities. A more detailed
comparison with experimental data and derivation of bounds will
appear in a forthcoming publication. The main results of our work
was the exponential decoherence-induced  damping in the oscillation
probability, which for the case of Gaussian fluctuations, with
variance $\sigma^2$, is given by (\ref{gauss2}), (\ref{ddef}), while
for the case of Cauchy-Lorentz distribution by (\ref{cl}).

For realistic models of quantum gravity, we first observe that in
the Gaussian model, the damping exponent is much more suppressed
than the corresponding one in the Cauchy-Lorentz
(Lindblad-like-time-scaling) case.  This is due to the extra
suppression factor $(m_1^2 - m_2^2)/E$ appearing in the exponent of
the Gaussian model. In order to make direct comparison with
experiment, it is essential to use as concrete quantum gravity
models as possible. To this end, the D-particle foam
model~\cite{horizons} turns out to be very useful. According to this
model, the stochastic metric fluctuations are \emph{induced}, as
explained in the introduction, by means of the recoil of the
D-particle defect during its topologically non-trivial
interaction/capture process with matter. Considering a gas of
D-particles, it is natural to consider a Gaussian model for the
distribution of the respective (spatial) recoil velocities, $u_i$,
which in turn induce the metric fluctuations (\ref{metric}), for
heavy defects. In such a case, the distribution variables are the
dimensionless ratios $u_i/c$, which however are restricted to be
less than one in magnitude, as $u_i$ are not allowed to exceed the
speed of light in vacuum $c$, as requested by string theory. In such
a case, the relevant integrals in the previous sections, e.g.
(\ref{cfgd}), have to be understood to be restricted to the range
$[-1,1]$.

In this sense, the above results can only be viewed as an
idealisation of the situation characterising the D-foam model, which
however, is a pretty good approximation of reality if the variance
of the Gaussian distribution is pretty small, which is the case in
the D-foam model. Indeed, taking into account that $\sigma^2$ in
such a case might \emph{naturally} be expected to have an  order of
magnitude dictated by the square of (\ref{velmag}), since after all
$\sigma^2$ describes the variance of D-particle recoil velocities,
which was assumed small, and the typical order of the latter is
given by (\ref{velmag}). If one assumes that the momentum transfer
of a particle probe, during its capture by a recoiling D-particle
spacetime defect, is of the order of the particles momentum itself,
then we may have for the variance $\sigma^2$ the estimate:
\begin{equation}
\sigma^2 = {\mathcal O}\left(\frac{E^2}{M_s^2}g_s^2\right)
\end{equation}
This yields a damping exponent (\ref{gauss2}) of order (in units $\hbar = c = 1$):
\begin{equation}\label{gauss3}
{\rm exp}\left(-\Omega^2_{\rm Gauss} t^2\right)~, \qquad
\Omega^2_{\rm Gauss} =  \frac{g_s^2 (m_i^2 - m_j^2)^2}{32~M_s^2}~.
\end{equation}

For the case of Cauchy Lorentz distribution of D-particle velocities
(adopted appropriately, as in the Gaussian case, to incorporate
velocity variables $u_i$ which do not exceed the speed of light in
vacuum $c$), one observes that the parameter $\gamma$ provides a
characteristic scale for D-particle velocities, since, if they
exceed that scale the distribution is diminished significantly. On
account of (\ref{velmag}), this leads to the assumption that a
natural order of magnitude for $\gamma$, in the context of the
D-particle foam model, is $\gamma \sim g_s E/M_s $, with $E$ the
energy of the particle probe. The corresponding decoherence-induced
damping exponent has then the form:
\begin{equation}\label{CL2}
{\rm exp}\left(-\Omega_{\rm CL} t\right)~, \qquad \Omega_{\rm CL} =  \frac{g_s |m_i^2 - m_j^2|}{2~ M_s}~.
\end{equation}

Searches for Lindblad decoherence~\cite{lisi1}, using the latest
neutrino experimental data, have bounded the respective coefficients
in a stringent way. For two-flavour models, the parametrization used
in \cite{lisi1} for the decoherence-induced Lindblad-type  damping
coefficients is:
\begin{equation}
{\rm exp}\left(-{\tilde \gamma}~t\right)~, \qquad {\tilde \gamma} = \gamma_0 \left(\frac{E}{\rm GeV}\right)^n
\label{lisi2}
\end{equation}
with the following bounds provided by means of combining atmospheric, solar-neutrino and KamLand data~\cite{lisi1}
\begin{eqnarray}\label{lisi1b}
\gamma_0 & < & 0.67 \times 10^{-24} ~{\rm GeV}~, ~~ n =  0\nonumber \\
\gamma_0 & < & 0.47 \times 10^{-20} ~{\rm GeV}~, ~~ n =  2\nonumber \\
\gamma_0 & < & 0.78 \times 10^{-26} ~{\rm GeV}~, ~~ n  =  -1
\end{eqnarray}
It should be remarked that all these bounds should be taken with a
grain of salt, since there is no guarantee that in a theory of
quantum gravity $\gamma_0$ should be the same in all channels, or
that the functional dependence of the decoherence coefficients
$\gamma$ on the probe's energy $E$  follows a simple power law.
Complicated functional dependencies $\gamma (E)$ might be present in
general.

To compare with our model above (\ref{CL2}), we observe that the
resulting damping coefficient is independent of the probe's energy
$E$, as a result of the $E$-dependence of the coefficient $\gamma$.
For such constant decoherence coefficients, the analysis of
\cite{lisi1}, (\ref{lisi1b}), yields the following bound on
$M_s/g_s$:
\begin{equation}\label{lisi3}
\frac{M_s}{g_s} > 0.74 \times 10^{24} ~{\rm max}_{i,j}\left[\left(\frac{|m_i^2 - m_j^2|}{{\rm GeV}^2}\right) ~{\rm GeV}\right]
\end{equation}
where max$_{i,j}$ indicates the maximum mass difference among
neutrino flavours, and we assumed that the quantum gravity
parameters are the same for all flavours, which is certainly the
case of the D-particle foam model. Recent data~\cite{lisi1} indicate
that $\Delta m^2_{ij} \in (10^{-23}-10^{-21})~{\rm GeV}^2$, from
which (\ref{lisi3}) implies:
\begin{equation}
\frac{M_s}{g_s} > 740~{\rm GeV}
\end{equation}
The expected minimal value of the string mass scale $M_s$ is, of
course at least a few TeV, for which the string coupling must be
very weak in order to provide realistic string phenomenology. For
couplings of order $g_s \le 1/2$ the phenomenologically correct
scale is close to four-dimensional Planck scale, $M_s \sim 10^{18}$
GeV. The above considerations, therefore, imply that the currently
available neutrino data do not have the sensitivity to probe
realistic D-particle foam models (the Gaussian Models decoherence is
much more suppressed than the Cauchy-Lorentz one, as already
mentioned).

One may reverse the logic, and try to consider bounds in as much
model independent way as possible, using the results (\ref{gauss2}),
(\ref{cl}), without reference to any explicit model for the
parameters $\sigma^2$ and $\gamma$. In such a case, one may assume
that these parameters are probe-energy independent, and try to bound
their values, by comparing with data. We postpone such a complete
analysis for a forthcoming work. Here, however, we note in brief
that in the Cauchy-Lorentz case with constant scale-parameter
$\gamma$, the resulting decoherence coefficient corresponds to the
$1/E$-dependent case, $n=-1$ in (\ref{lisi2}), for which, on account
of (\ref{lisi1b}), one obtains the following bound on $\gamma$:
\begin{equation}
     \gamma < 10^{-5}
\label{gammadistr}
\end{equation}
on account of the above-mentioned measured neutrino mass differences.

A final comment, concerns the order of the damping exponents in
(\ref{gauss2}) and (\ref{cl}) for astrophysical neutrinos. To answer
such a question it is imperative to know the energy dependence of
the respective damping coefficient. Since the decoherence effects
depend on the actual time of flight of neutrino $t$, the effects are
maximised for extraterrestrial neutrinos, coming from extragalactic
sources. Potential limits on Lindblad decoherence using high energy
(TeV scale) astrophysical (anti)neutrinos from ICE-CUBE in the
future have been analysed in \cite{icecube} and also in
\cite{highnu}. In the ICE-CUBE case \cite{icecube}, the inverse
energy $n=-1$ decoherence coefficient $\gamma_0$ in the terminology
(\ref{lisi2}), is found to be (at a 90\% CL) $\gamma_0 <
10^{-34}$~GeV, which will improve the existing sensitivity by eight
orders of magnitude, implying, for a constant Cauchy-Lorentz
distribution, the limit $\gamma < 10^{-13}$. Such sensitivities are
not far from naturally expected values of $\gamma $ in space-time
foam models involving heavy (Planck-mass $M_s/g_s \sim 10^{19}$ GeV)
D-particles and TeV-momentum transfers during the capture of
TeV-energy (anti)neutrino matter by the recoiling D-particle.
Moreover, for completeness, we mention that, as discussed in
\cite{icecube}, the 90\% CL bound on the inverse-energy decoherence
from ICE-CUBE will imply, according to (\ref{lifetime}) the existing
bounds on the electron-antineutrino life time $\tau_{{\overline
\nu}_{\rm e}}/m_{{\overline \nu}_{\rm e}} > 10^{34}$~GeV$^{-2}$,
improving by four orders of magnitude the existing bounds from solar
neutrinos.

Finally, we mention that in \cite{waxman} it was argued that
ultra-high energy neutrinos, with energies $10^{17}-10^{19}$ eV can
be emitted by Gamma Ray Bursts (GRB), which actually carry a
significant fraction of the GRB energy. Most of GRBs lie at
cosmological distances corresponding to distances $z > 1$, i.e.
larger than $10^{27}$~m. To study self-consistently such cases, one
needs of course to consider the propagation of the neutrino in a
Robertson-Walker background, about which one could expand the
space-time metric fluctuations. We hope to come to a study of such
issues in a future work. However, to get a preliminary idea it
suffices to ignore the expansion of the Universe and consider the
Cauchy-Lorentz decoherence distribution as a pilot case. For such
distances and energies, the relevant exponent (\ref{cl}) of a
Cauchy-Lorentz decoherence model with constant
(probe-energy--independent) $\gamma < 10^{-5}$ (c.f.
(\ref{gammadistr})), becomes roughly of order: $5 \times 10^{-37}
\times t $, where we took into account that the maximal of the
neutrino mass difference is ${\rm max}_{i,j}\Delta m_{ij} \sim
10^{-3}$. Such damping becomes of order one for distances of order
$L \sim 10^{27}$~m, i.e. at cosmological distances of GRBs. Of
course, the expansion of the Universe, will modify such results,
nevertheless what this preliminary exercise showed is that high
energy astrophysical neutrinos can indeed constitute sensitive
probes of decoherence models~\cite{icecube,highnu}. We hope to be
able to come back to a detailed discussion of such issues in the
future.

\section*{Acknowledgements}

The work of N.E.M is partially supported by the European Union through the FP6 Marie
Curie Research and Training Network \emph{UniverseNet} (MRTN-CT-2006-035863).

\end{document}